# Machine Learning with DBOS

Robert Redmond*
MIT

Nathan W. Weckwerth*
MIT

Brian S. Xia*
MIT

Qian Li
Stanford University

Peter Kraft
Stanford University

Deeptaanshu Kumar
CMU

Çağatay Demiralp
Sigma Computing

Michael Stonebraker
MIT

## ABSTRACT

We recently proposed a new cluster operating system stack, DBOS, centered on a DBMS. DBOS enables unique support for ML applications by encapsulating ML code within stored procedures, centralizing ancillary ML data, providing security built into the underlying DBMS, co-locating ML code and data, and tracking data and workflow provenance.

Here we demonstrate a subset of these benefits around two ML applications. We first show that image classification and object detection models using GPUs can be served as DBOS stored procedures with performance competitive to existing systems. We then present a 1D CNN trained to detect anomalies in HTTP requests on DBOS-backed web services, achieving SOTA results. We use this model to develop an interactive anomaly detection system and evaluate it through qualitative user feedback, demonstrating its usefulness as a proof of concept for future work to develop learned real-time security services on top of DBOS.

## 1 INTRODUCTION

We earlier introduced the database-oriented operating system (DBOS), a new operating system stack [37] that stores all system and application state in database tables and executes the operations on state as transactions. We have been building and experimenting with it in phases since then. DBOS offers several benefits over traditional OSes, which we discuss in detail elsewhere [23, 37]. Some of the DBOS benefits are directly relevant to machine learning (ML). In this paper, we first discuss these benefits in general. We then demonstrate the practicability and scalability for a cross-section of them through two application cases: (1) Serving GPU-accelerated deep ML models on DBOS, where programs are implemented and executed as stored procedures, and (2) developing an end-to-end anomaly detection service, where the driving ML model and the affordances of the interactive visual analysis interface leverage provenance data for training and contextualization, respectively.

ML has become ubiquitous with applications across domains [4] from molecular biology to law. ML models are increasingly used to automate and augment software systems and user tasks, causing many applications to be redesigned [19, 43]. ML is, however, typically built on top of most systems as secondary services or applications. It is often difficult for developers to find adequate training data, address privacy and scalability concerns, or track and manage updates to data as well as models. Application and system software would benefit from an OS stack that provides first-class support for ML development and deployment. DBOS offers several advantages in this context.

**Data and Code Together** While most existing cloud systems disaggregate compute and data, DBOS tightly integrates them, co-designing its function execution (stored procedure) subsystem with a distributed DBMS. Since ML development and serving are data-intensive, DBOS provides significant performance gains for data-intensive operations by bringing data and code together [23]. Its serverless environment reduces latencies due to data transfers over network, and simplifies task scheduling and memory management, enabling performances exceeding those provided by Amazon Lambda [1] and Open Whisk [9].

**Security and Privacy** With the increasing importance of data protection and privacy requirements (e.g., GDPR and HIPAA), challenges in sandboxing model training and serving have become a rate-limiting factor in adopting ML in the current multitenant cloud computing settings. As a result of the above design choice, co-locating data and code, DBOS executes all user applications and OS services as stored procedures. This in turn provides a time-tested means for sandboxing ML development and serving, benefiting from the built-in access control mechanisms of the underlying DBMS. DBOS's program execution model also offers an effective solution for training and serving ML models without transferring users' data outside the database. This preempts and simplifies a multitude of privacy and security concerns along with associated contractual hurdles in ML model development and serving in enterprises. We can further adopt recent work [2, 14] on database security and privacy into DBOS.

**Serverless Computation Workflow** DBOS provides a function-as-a-service, or serverless, model of computation [23] where users write large programs as graphs of smaller functions and submit these to a remote service for execution. Serverless computing is becoming popular because it enables transparent application autoscaling, dramatically reducing the complexity of managing cloud services [18]. We believe a serverless computing model will greatly benefit ML development and deployment, which often must run at scale in distributed settings.

**Data Governance and Observability** Developing and deploying ML models is an iterative process based on trial and error. ML developers experiment with new datasets, models, and parameters to achieve their modeling goals. ML models also often need to be deployed and regularly updated to improve performance or preserve them against shifts in data distributions that may occur in application domains. DBOS makes model management [27, 40, 45] easier by centralizing and tracking relevant model data and parameters. Since model performance is often determined by the quality and the size of training data, reproducing various versions or checkpoints in this process is critical for debugging. The data and workflow provenance of DBOS offers opportunities to improve the ML development (e.g., training, debugging, introspection, etc.) and serving experience. DBOS augments data provenance information with workflow provenance to further facilitate interpretability. It records executions of user functions along with related metadata and associates them with data provenance information. Moreover, since we co-locate data and compute, we can potentially leverage ideas from prior work [28] to automatically capture fine-grained data provenance.

**Support for Heterogeneous Hardware** Hardware like GPUs, TPUs, SSDs, and FPGAs have become omnipresent in compute clusters, bringing constraints as well as optimization opportunities for ML and other data-centric applications at scale. Performant ML models heavily rely on heterogeneous hardware such as GPUs and TPUs. DBOS proposes directly supporting and managing such hardware through its stored procedure interface freeing users from the headaches of provisioning.

**Training Data** Given the current state of ML technology centered on high-capacity models, the availability of large-scale training data is perhaps the most important requirement for wider and deeper adoption of ML. The data generated by DBOS provenance tracking can facilitate the integration of ML to improve core OS components as well as applications built on top of it.

In the following, we present results from two of our ongoing ML-related projects around DBOS. In the first project (Section 2), we investigate the performance of serving ML models in DBOS stored procedures while using heterogeneous hardware such as GPUs. We show that image classification and object detection models using GPUs can be served as DBOS stored procedures at scale without incurring significant performance loss. In the second project (Section 3), we demonstrate how provenance data collected by DBOS can be used to develop ML-driven applications to augment system services. To that end, we present an ML model trained on HTTP request logs collected by a DBOS-backed web service called Nectar Network[1]. Using byte pair encoding along with a 1D CNN, we achieve state-of-the-art (SOTA) performance. We use this model to drive an interactive anomaly detection system and demonstrate its usefulness through the qualitative feedback we collect.

---

[1]https://github.com/DBOS-project/apiary/tree/main/postgres-demo

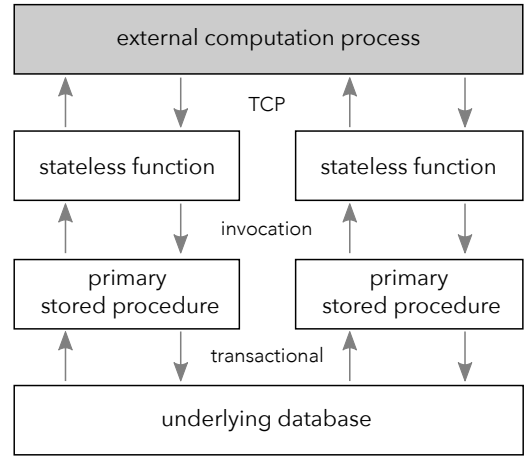

**Figure 1: DBOS architecture for using external services. Upward lines represent the flow of data originating from the underlying database. Downward lines represent the flow of information computed by an external process.**

## 2 HETEROGENEOUS HARDWARE SUPPORT

DBOS proposes built-in support for heterogeneous hardware in a serverless computation platform with tightly integrated computation and storage layers. Combined with data and workflow provenance tracking, this makes DBOS compelling for developers to use in data- and compute-intensive tasks, including machine learning development and deployment. DBOS can run compute-intensive tasks as external services and thus utilize specialized hardware such as GPUs. Below we briefly describe the asynchronous programming model architecture of DBOS and then results from ML models we deployed using it.

### 2.1 System Architecture

In DBOS, programmers develop applications as a collection of stored procedures, which use embedded SQL queries to communicate with the underlying SQL DBMS (VoltDB [38] in our case) instance. These stored procedures run natively as ACID transactions in the database. VoltDB achieves high OLTP performance in part by a single-threaded-run-to-completion stored procedure execution model. However, this model is ill-suited for computationally intense tasks such as ML. To circumvent this problem, DBOS enables developers to asynchronously invoke stateless functions that are not tied to the database. These stateless functions connect via TCP to an external process that manages the computation. Figure 1 shows the high-level architecture of DBOS's asynchronous programming model.

The entry point for an end-user to invoke a compute-intensive task is a single primary stored procedure. This stored procedure retrieves data from the database and asynchronously invokes a stateless function, and the primary stored procedure also asynchronously invokes another stored procedure to handle the return value from the stateless function. This stateless function does not have access to the database and thus can be run concurrently with other stored procedures. It connects over TCP to a long-running external process managing computation. After computation occurs, the computed information is transferred back to the other stored procedure, at which point it may be inserted into the database. Multiple stored procedures may be invoked, leading to multiple stateless functions making TCP connections with the external process.

| task ID | piece ID | data |
|---|---|---|

**Figure 2: Structure of a message sent to the external process. Each message includes an identifier denoting which task the message belongs to and, if necessary, an identifier resolving which piece of the puzzle the message represents when pooling data from multiple stateless functions.**

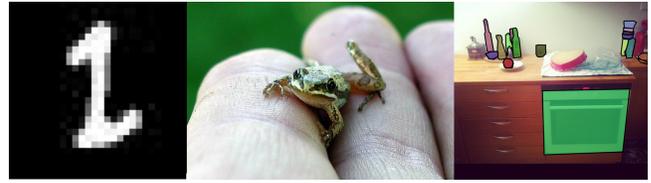

**Figure 3: Example images from each dataset from left to right: MNIST, ImageNet, and COCO.**

**Primary Stored Procedure and Stateless Function** A user-defined end-to-end workflow performs the following operations: (1) invokes a stored procedure to fetch the desired data from the database; (2) asynchronously invokes a stateless function and gives it the retrieved data. When the stateless function receives a response from the external process managing computation, it returns the computed information; (3) asynchronously invokes another stored procedure to insert the computed information into the database. Note that operations (1) and (3), which directly interact with the database, run as ACID transactions.

Even for a single computational task, data may need to be transferred using multiple invocations of the primary stored procedure. This is due to the fundamental size limitations of VoltDB as a main-memory database. VoltDB restricts the maximum column and row sizes to respectively be 1MB and 2MB. Similarly, the maximum return size of stored procedures is 50MB. Moreover, the current DBOS implementation restricts stored procedures to 1MB due to the way it serializes tasks. Regardless of particular size limits, one can easily imagine computational tasks requiring significantly more data than what can be transferred at a single invocation. Nonetheless, this is not an issue since the external process will accept as many connections as needed.

**External Process** An external process is a long-running process that accepts TCP connections from potentially many stateless functions. This has several advantages over doing computation directly in stateless functions: (1) The external process can initialize computation structures (e.g., loading saved ML models) before receiving any connections, eliminating potential long delays in computation. (2) The limitation on the amount of information returned by a stored procedure means that data from multiple stateless functions may need to be pooled to run a single computation. (3) Transferring data over TCP means the external process may be run in any high-level language, not just Java, and may potentially be run on any machine. Figure 2 shows the structure of a message sent to the external process. The external process determines which task the data is for and how to assemble multiple pieces of data based on the identifiers at the start of the message.

The architecture we have described thus far is quite general, and one can envision using this architecture to run many types of compute-centric tasks. It also provides several advantages for ML model deployment: (1) Data sent to the external process can easily be used for either inference or training the model itself, by simply changing the task ID referenced in Figure 2 to tell the external process which purpose to use it for. (2) Inference and training can be performed immediately by the external process, or it can wait until a certain batch size appears. This allows users to balance work done by the external process between different tasks or prioritize a specific task over others. (3) The models themselves can be stored in DBOS and loaded into the external process by simply storing and sending 1MB chunks of the serialized model. Combined with DBOS's robust provenance system, this allows users to easily track not only which data was used for training and inference, but also the history of the model itself as it is trained on new data or experiences architectural changes from the end user.

In our current implementation, we assume each compute intensive task can run in a single external process. However, it is possible to extend our system to interact with a group of external processes for supporting distributed tasks, for example, large-scale ML models (e.g., OpenAI GPT-3) inferences.

### 2.2 Evaluation

Now we evaluate this architecture for serving image classification models.

**Datasets and Models** We select three popular image classification datasets with images at different scales: MNIST [22], ImageNet [7], and COCO [25]. Figure 3 shows an example from each dataset. We believe that our datasets and corresponding models represent realistic use cases for a user deploying an image classification model.

The MNIST dataset consists of 70,000 images of handwritten digits. Each image is black and white and has a fairly small size of 28×28 pixels. The MNIST model we use has a single hidden layer with 512 nodes (or units). This is in line with the scale of other MNIST models [22].

The ImageNet dataset [7] is an image dataset organized by the WordNet hierarchy [39], containing over 14M images. The images in ImageNet are scrapped from various sources on the web and vary significantly in size, with an average size of around 470×390 pixels—most models often resize the images to a standard 256×256 or 224×224 pixels before using them. Training a model to perform with high accuracy on the ImageNet dataset can be quite costly compared to training a classifier for the MNIST dataset. Since developing novel or extremely accurate models is not the goal of our evaluation, we consider only the standard 1,000 categories that are used by the ILSVRC challenge [34], an annual image classification competition run by the curators of ImageNet. More importantly, we perform transfer learning using a pre-trained headless model [15]. The pre-trained model uses the MobileNetV2 [35] architecture and was originally trained on the ImageNet dataset. It takes a 224×224-pixel image as input and outputs a feature vector with 1,280 features; on top of this, we add a layer acting as the classification head, mapping these features to the 1,000 categories from ILSVRC. This approach saves significant time on training while allowing the model to perform quite well.

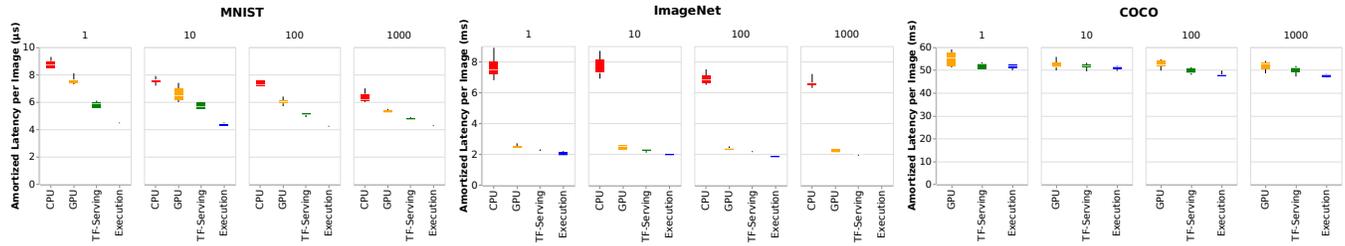

**Figure 4: Comparing the external process on a CPU and GPU against TensorFlow-Serving and the baseline time spent executing the model for testing across (from left to right) MNIST, ImageNet, and COCO datasets.**

The COCO (Common Objects in Context) dataset [25] contains 328,000 images of 91 distinct and easily recognizable object types such as "clock", "horse", and "traffic light". Not only is each image labeled with each object type found in it, but each individual object is marked with a bounding shape. The images have an average resolution of around 640×480 pixels, and models typically resize them to standard sizes such as 512×512 pixels before using them. Similar to the ImageNet dataset, there are high costs in training a model to perform with high accuracy. Thus for simplicity, we directly use a pre-trained model provided by TensorFlow on TensorFlow Hub. The model [16] uses the CenterNet [47] architecture with the Hourglass [29] backbone and was originally trained on the COCO dataset. It takes a 512×512-pixel image as input and outputs data identifying the objects in the images and bounding boxes for each object.

**Procedure and Setup** For performing inference, each invocation of the primary stored procedure fetches as many images as can comfortably fit in 1MB. In all three datasets, the sizes we have chosen to scale the images to are less than 1MB, so there is no need to break an image into multiple pieces in the database.

The external process runs on a virtual machine equipped with a GPU configured using Google Cloud Platform [31] and uses the GPU to perform inference. To measure inference performance, we provide comparisons against running on a CPU and against the time spent directly running the model, as well as a comparison against TensorFlow-Serving [30], a popular machine learning deployment system. We anticipate some overhead on directly running the model, but even against TensorFlow-Serving, there will be some overhead incurred primarily by invoking stored procedures. For each comparison we record performance numbers for 1, 10, 100, and 1,000 invocations of the primary stored procedure.

### 2.3 Results

Table 1 and Figure 4 respectively show the inference (testing) overhead and amortized latencies across datasets. The results are promising, with overheads of less than 10% when compared with TensorFlow-Serving for both the ImageNet and COCO datasets, which contain significantly larger images than the MNIST dataset. Even for the MNIST dataset, with enough invocations of the primary stored procedure the overhead drops to just over 10%, and a single invocation yields performance just under 30% worse than that of TensorFlow-Serving. In general, the performance hit incurred by deploying machine learning models using DBOS is negligible compared to the cost of computation. We discuss the results per model in more detail below.

While simple, our MNIST model achieved 96% accuracy or more on the testing set. Since the MNIST images are only 28×28 pixels and they are black and white, each only takes up 784 bytes, indicating 1,200 images could comfortably fit into a single MB and thus we fetch this many with each invocation of the primary stored procedure. The external process simply returns the classification of each image. We pay a relatively high overhead compared to both model execution time and TensorFlow-Serving, especially when only invoking a single stored procedure. We can also see in Figure 4 that the GPU is not dramatically better than the CPU for this example. Both of these facts are understandable as the computation here is relatively simple compared to the other inference tasks, with inference taking only a few microseconds for each image. Even so, the overhead is not too large; with 1,000 invocations of the primary stored procedure the overhead against TensorFlow-Serving falls to just over 10% and the overhead against the base model execution time is 23.8%.

The ImageNet images have been scaled to 224×224 pixels and they are color images, suggesting 6 images can comfortably fit into 1MB. Thus we fetch 6 images with each invocation of the primary stored procedure. The external process simply returns the classification of each image. Here the overhead against model execution time and TensorFlow-Serving is, unsurprisingly, drastically better than for the MNIST example. Note that for this model the GPU outperforms the CPU by a factor of 3 (Figure 4) due to the computation being more complex. For our system using DBOS on a GPU, even when invoking a single stored procedure the overhead against TensorFlow-Serving is less than 10%, dropping to just 5% at 1,000 invocations.

The COCO images have been scaled to 512×512 pixels and they are color images, suggesting only a single image can fit within 1MB. Thus each invocation of the primary stored procedure fetches a single image. For this example the external process returns more than just a simple classification for each image; instead, it returns data corresponding to multiple objects identified in the image and estimated bounding boxes for each. Functionally this does not change the implementation much as the information is still just sent back to the stateless function. Here the overhead against model execution time and TensorFlow-Serving is better than both of the previous examples. Figure 4 shows the performance numbers in more detail, although the CPU numbers are omitted as for this test the CPU is orders of magnitude slower due to the complexity of the computation. In all cases for this model the overhead incurred is less than 10%, even when comparing directly against the model execution time, and the overhead against TensorFlow-Serving is 5% or less.

|  | 1 Invocation | | 1,000 Invocations | |
| --- | --- | --- | --- | --- |
|  | TF-Serving | Execution | TF-Serving | Execution |
| MNIST | 28.8% | 72.7% | 10.6% | 23.8% |
| ImageNet | 9.1% | 26.3% | 5.0% | 16.7% |
| COCO | 5.1% | 6.7% | 3.3% | 9.8% |

Table 1: Overhead incurred by DBOS with the external process running on a GPU in comparison to TensorFlow-Serving and raw model execution time across MNIST, ImageNet, and COCO datasets. The overhead decreases at scale.

## 3 ANOMALY DETECTION

DBOS tracks provenance for all VoltDB queries and updates. This includes all OS state, any user state stored in VoltDB, and all stored procedure invocations. The built-in data collection of DBOS brings many opportunities for data-driven approaches for monitoring and debugging. In particular, it facilitates ML applications for system services, providing plentiful training data. Monitoring and preventing anomalous activity are key for security applications. Here we introduce a new ML model trained on provenance data to detect anomalous HTTP requests in real-time. This model provides high prediction accuracy, achieving state-of-the-art results. We then present the user interface of an interactive system built on this model. We evaluate the user interface with two qualitative user studies, one formative (initial) and one summative (final). Results suggest substantial improvement in the usefulness of the user interface over its iterations informed by the formative study. Crucially, our interactive tool serves as a proof-of-concept for future work to develop end-to-end security analysis system services or applications on DBOS.

**Provenance Capture** We designed the DBOS stack to support robust data and workflow provenance. A high-performance DBMS can store structured, provenance information that is easily accessible through SQL queries. In practice, VoltDB [38] serves as the main memory DBMS that stores OS state. However, VoltDB is unsuited for handling large amounts of provenance data, requiring data to be spooled to Vertica [21] for long-term storage and downstream analysis. Vertica significantly outperforms VoltDB in performance with regard to provenance queries on tables that have larger than $10^5$ rows [20]. This clearly highlights the importance of a dedicated OLAP system like Vertica to serve as the storehouse for provenance data.

One important question is what can be answered through such a data provenance system. Below are some potential queries of interest.

- **Table history**. Who was the last person to write to a particular table? Which table had the most updates over an arbitrary time frame?
- **Compromised users**. What are all of the blocks that were read or written by a compromised user over an arbitrary time frame? Who are all of the users that read a compromised block over an arbitrary time frame?
- **Chain of provenance**. What are all of the blocks that may have resulted from reading a particular block (downstream)? What are all of the blocks that may have influenced a particular block (upstream)?
- **Debugging**. What is the exact state of a table at a particular point in time?

**Web Application Attacks** Web applications have quickly become one of the most popular platforms for information and service delivery [6, 24]. They have several features that have led to their success, such as remote accessibility, cross-platform compatibility, and fast development. As a result, web applications are also used for providing services such as healthcare and financial services that often handle sensitive data. On the other hand, web applications are the most common attack vector (a means by which an attacker can gain access to a network server) used for intrusion, resulting in the most breaches and compromised data [3]. In the following sections, we detail the most common types of web application attacks, including the methodology and end goal.

*SQL injection* A SQL injection attack occurs when a malicious user tampers with the SQL queries sent by the web application to its corresponding database [8, 24]. This occurs when SQL keywords or operators are inserted into queries without input sanitization to explicitly remove or filter them out. This can be done through malevolent insertions into user inputs (e.g. fillable fields), cookies, and/or HTTP headers. Importantly, the contents of the insertion dictate whether the attack is of the first- or second-order. First-order attacks are executed immediately with the intent to return results immediately. In other words, the entire attack is localized within the insertion. A concrete example is using the union keyword to attach malicious SQL queries to the end of standard SQL queries. Second-order attacks rely on an initial insertion that lies dormant for some period of time, usually until a follow-up insertion prompts the execution of the first insertion. A concrete example is inserting an initial malicious query that can be prompted at a later date. The follow-up query would return metadata on users who have accessed the web application since the initial insertion. The overall purpose of these attacks may be to steal credentials, alter data, delete data, and/or access connected resources.

*Cross-site scripting* Cross-site scripting (XSS) occurs when a malicious user is able to execute custom scripts in a victim's browser [24]. This typically occurs when web responses are unsanitized, meaning they are unchecked for special characters/keywords that may lead to unexpected or malicious behavior. This becomes problematic when web applications utilize the same-origin policy, which allows scripts in a webpage to access the data in another webpage if they

```
{
  "getAuthType" : null,
  "getPathInfo" : null,
  "cookie: JSESSIONID" : "D8B035F1CD70BC14F7E1804C54911F2B",
  "getRemoteAddr" : "171.66.11.171",
  "getServletPath" : "/home",
  "getMethod" : "GET",
  "getContextPath" : "",
  "getServerName" : "nectarnetwork.org",
  "getPathTranslated" : null,
  "getRequestedSessionId" : "D8B035F1CD70BC14F7E1804C54911F2B",
  "header: user-agent" : "Mozilla/5.0 (Macintosh; Intel Mac OS X 10
  "getRequestURL" : "http://nectarnetwork.org/home",
  "header: upgrade-insecure-requests" : "1",
  "header: host" : "nectarnetwork.org",
```

**Figure 5: Snippet of an HTTP request log from Nectar Network.**

both come from the same origin (combination of URI scheme, host name, and port number). For instance, an attacker could insert a malicious script into a less secure webpage in order to access confidential information from a more secure webpage. Similar to SQL injection, there are first- and second-order attacks that dictate the timing of when the attack occurs. A first-order attack, such as reflected XSS, prompts the user to click on a custom link which delivers an XSS payload to the web application. This payload allows the attacker to perform any action that the user would be able to perform. A second-order attack, such as persistent XSS, may rely on sending the XSS payload to a back-end database (e.g. through usernames, comments, forum posts, etc.) that gets triggered once a victim loads a webpage containing the relevant information. These attacks are often used to steal sensitive information about a victim such as credit card information, medical records, and/or cookie details.

*Distributed denial-of-service* Distributed denial-of-service (DDoS) occurs when a malicious user overwhelms a target resource with superfluous traffic, rendering the resource unable to respond to legitimate traffic in a timely manner [33]. It should be noted that the superfluous traffic comes from a wide variety of sources (i.e. the "distributed" aspect), which makes it much more difficult to differentiate and block the multiple sources of such traffic. This attack is not specific to web applications, but it remains one of the most common attack patterns due to its generality and effectiveness. The primary purpose of a DDoS attack is to render a web application inoperable, thereby disrupting its normal function and inconveniencing its users. Some secondary purposes that directly result from a DDoS attack include extortion, reputational damage, and/or financial drain.

**Nectar Network** Nectar Network is a simple web application developed on top of DBOS. It serves as a rudimentary social networking site and is publicly accessible at nectarnetwork.org. We made Nectar Network publicly available in order to capture real-world internet traffic, thereby allowing us to test DBOS provenance capture and develop real-time anomaly detection using a realistic web application deployment. All HTTP requests are logged and stored in Vertica [21] using the fields shown in Figure 5. This schema loosely follows the W3C extended logging format as described by Microsoft [44]. The format contains enough information to form a complete history of an HTTP request.

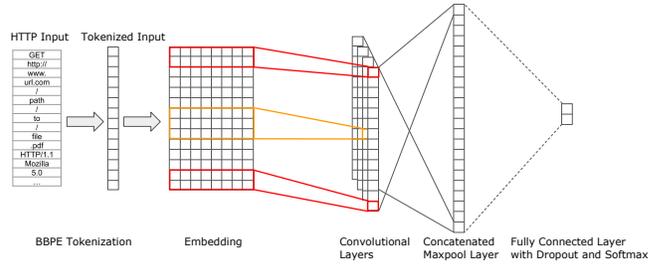

**Figure 6: BBPE CNN model architecture.**

### 3.1 Model

Our model (Figure 6) has two basic components: tokenization and classification. The tokenization component uses byte-level byte pair encoding (BBPE) to break down the input bytes into byte tokens that hold semantic meaning. In the classification step, a convolutional neural network (CNN) takes the token bytes as input and outputs the predicted probability of anomaly.

**Byte-level Byte Pair Encoding** Byte pair encoding (BPE) was introduced as a method of compressing strings [10]. The technique uses the characters of a string as tokens, and additionally adds tokens representing the most common combinations of characters present in a string. By doing so BPE is able to outperform Lempel–Ziv–Welch compression in terms of compressed data size at the cost of increased time for compression. Note that after tokenization, there is a dictionary mapping seen characters to assigned tokens known commonly as a vocabulary. Because BPE can tokenize a string without loss of information, it can serve as a tokenizer for language machine learning techniques [36]. The authors report that this tokenization method serves well for vocabularies in which there are very rare words and words that are out of vocabulary. Byte-level byte pair encoding (BBPE) is a tokenization method that builds on BPE but operates on bytes instead of characters [42]. This is particularly powerful because BBPE guarantees that there will be no unknown tokens. In the worst case, an input can be tokenized as its individual bytes, meaning unique characters that have not been seen before can still be tokenized. Since HTTP requests and other machine code often includes unique characters, and in particular injection attacks use unique characters to confuse web application, this characteristic makes BBPE a strong choice for tokenizing our HTTP requests.

Incoming HTTP requests are formatted into a single string which includes all of the fields separated by spaces. An example string is "`GET http://url.com/path HTTP/1.1 [User-Agent] [Content-Length] ...`", in which ... represents additional HTTP fields. We then collect and use these to train the BBPE tokenizer.

**Convolutional Neural Network (CNN)** Once BBPE tokenizes the HTTP request, the request is classified by a CNN model. The full architecture of the model is shown in Figure 6.

*Token embeddings* Earlier work [46] shows that learning embedded tokens as part of the CNN works well for task-specific text applications, such as detecting web attacks. This helps the learned embedding relate more closely with the desired classification, in this case, whether an HTTP request is malicious. This stands in contrast to embedding techniques such as Word2Vec, which aims to learn the semantic meaning of tokens. While this makes sense for actual languages, HTTP requests often lack this semantic meaning. While some tokens are words, others might simply be characters such as %20, the URL escape sequence for the space character.

Table 2: 5-fold cross-validation performance across datasets.

| Dataset | Accuracy | Precision | Recall | F1 | F1$_{std}$ |
|---|---|---|---|---|---|
| CSIC | 0.999 | 0.999 | 0.998 | 0.998 | 4.95×10$^{-4}$ |
| Sigma | 0.999 | 0.996 | 0.996 | 0.996 | 27.4×10$^{-4}$ |
| Nectar | 0.999 | 0.999 | 0.999 | 0.999 | 2.26×10$^{-4}$ |

*Convolution layers* For the convolutional layers, three kernel sizes (2, 3, 4) were used. For each of these kernel sizes, 100 filters were used reaching a total of 300 filters in this layer. Each of these layers uses the ReLU activation function, a piecewise linear function that returns positive values directly and negative values as zeroes. This generates the possible feature map of the HTTP request.

*Maxpooling layer* This layer takes the maximum of each feature map generated by the convolutional layers and concatenates them into a single vector.

*Dense layer* The dense layer fully connects the maxpool layer to the output layer to perform binary classification. In addition, to avoid overfitting during training, a dropout layer was included which zeroed out inputs with a probability of $p = 0.2$.

*Output layer* The output layer is a softmax layer with two nodes corresponding to the problem classes, turning the activations from the fully connected layer to class probabilities.

## 3.2 Model Evaluation

**Metrics** We report the precision, recall, and $F_1$ scores along with the accuracy for our model's performance across evaluation datasets and conditions. Note that the accuracy score alone provides an incomplete picture of performance, particularly when we have unbalanced class distributions in datasets. For example, in the Nectar Network dataset, a high proportion of events captured are malicious in nature. As such, a classifier could label all events as malicious and still achieve a high accuracy due to the low total number of non-malicious events.

**Datasets** We use three HTTP request datasets with different characteristics for our evaluation.

*CSIC* The CSIC dataset is a public benchmark [12], which earlier methods of anomaly detection often used to report results. This dataset consists of generated traffic, and so the labels are known to be accurate. We use this dataset to compare our mode with a select sample of earlier work.

*Sigma Computing* The CSIC dataset is automatically generated and fairly balanced. In real-world applications, this is rarely the case. The Sigma Computing dataset contains anonymized HTTP logs at Sigma Computing [5, 11], representing real-world activity. The logs were labeled by Cloudflare. The majority of the traffic logged is benign in nature. Less than 4% of the events logged are malicious.

Table 3: Performance of our model on the CSIC dataset with decreasing ratios of training data used in train-test splits.

| % Train | Accuracy | Precision | Recall | F1 |
|---|---|---|---|---|
| 80% | 0.998 | 0.997 | 0.997 | 0.997 |
| 50% | 0.995 | 0.994 | 0.995 | 0.994 |
| 25% | 0.985 | 0.969 | 0.994 | 0.982 |
| 10% | 0.957 | 0.907 | 0.996 | 0.949 |
| 5% | 0.953 | 0.946 | 0.938 | 0.942 |

*Nectar Network* This dataset consists of HTTP requests from Nectar Network provenance capture which was labeled using the "Registered" ruleset for Snort v2.9.19. This ruleset contains 43,091 rules, with any violations being logged and the corresponding entry being labeled as anomalous. Entries with no rule violations were considered to be benign. Since the website is available for public access, the majority of the traffic it generated was from webcrawlers or malicious users. As a result, the data is heavily weighted towards events labeled as malicious, with only about 7% of the traffic being benign.

**Baselines** We compare our model with three top-performing models selected from the literature: HTTP2vec [13], Code Level CNN [17], and SAE [41]. HTTP2vec also uses BBPE to tokenize the inputs, but it obtains the token embeddings via RoBERTa [26] and feeds them as input features to a support vector machine classifier. Code Level CNN uses a similar CNN architecture, but only tokenizes the inputs based on special characters. SAE uses n-grams before extracting features using a stacked autoencoder.

**Results** Table 2 shows the performance of our model measured through a 5-fold cross-validation on all of the datasets. The results reported are the averages over the 5 splits, along with the standard deviation of the F1 score. The model achieves high prediction performance across all metrics and datasets.

Table 3 depicts how the performance of the model changes with different amounts of training data. While performance does decrease, it's surprising just how well the model can do with only 5% (~3,000) examples to train on. This suggests that BBPE, which is trained on the entire dataset, is doing the heavy lifting for representation learning and making it easier for the CNN to learn a proper classifier. One possible instance of this lies in SQL injection attacks, which often obscure their intent by using URL encoding. %25 decodes as %, which is the URL escape character. This can be exploited against poorly secured web applications. Since BBPE operates at the byte level, it is able to recognize %25 as a unique token, which gives the CNN a very easy way to identify this as a feature of attacks.

Table 4: Performance of models on the CSIC dataset.

| Method | F1 |
|---|---|
| BBPE CNN (ours) | 0.998 |
| HTTP2vec [13] | 0.969 |
| Code Level CNN [17] | 0.963 |
| SAE [41] | 0.841 |

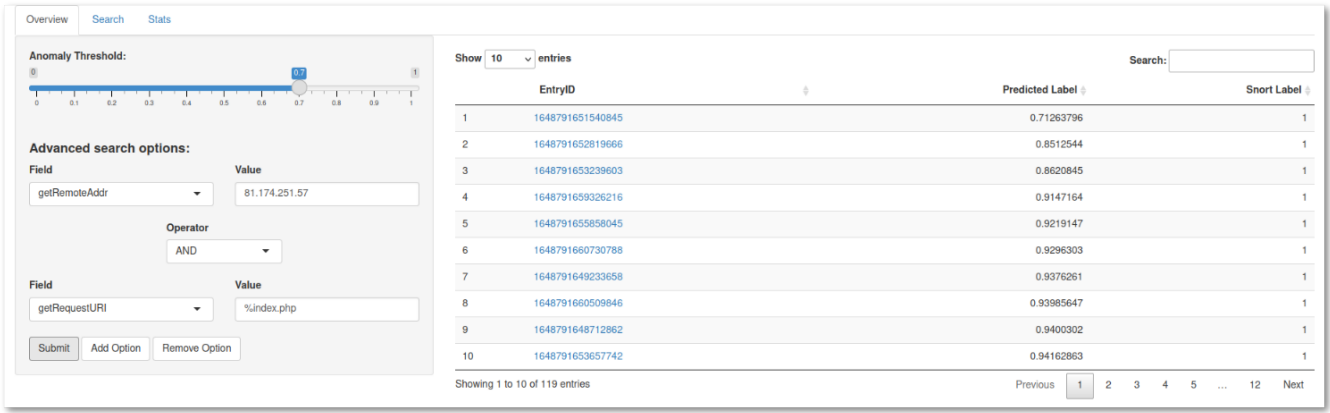

Figure 7: User interface has three main views (pages): Overview, Search, and Stats. Users can navigate to these views by clicking on the corresponding tabs (upper left). The default view is Overview.

Table 4 compares the performance of our model with the baselines on the CSIC dataset. Our model outperforms the best performing baseline by about 3% in F1 score.

## 3.3 User Interface

We develop an interactive visual analysis tool driven by our ML model above. Our goal is twofold: to help security administrators to further investigate predicted anomalies within the context of DBOS provenance data, and to demonstrate an end-to-end utilization of DBOS provenance tracking. We below focus on the user interface of the tool and its evaluation.

The user interface has three main views (or pages), which can be navigated through three corresponding tabs. The first tab (default tab on launch) is the overview page (Figure 7), which allows the user to apply various filters to the provenance data for interactive display. The second tab is the search page (Figure 9), which allows the user to directly query the provenance data using SQL commands as input. The third tab is the stats page (Figure 10), which allows the user to visualize anomaly trends through a historical line graph of detected anomalies.

**Prioritizing Anomalies** The goal of the overview page is to provide an interface for a user to filter anomalies based on the available logged fields and an anomaly threshold. This allows for anomaly prioritization based on the filtering options selected. An example of how this process is enacted can be seen in Figure 7. The left sidebar specifies the available filtering options and the right main content displays the output data table based on the selected filtering options.

The sidebar consists of two components; the anomaly threshold and advanced search options. The anomaly threshold is a slider between 0 to 1 that specifies the minimum value of the predicted label that should be displayed in the output data table. The advanced search view enables a user to easily filter for any number of HTTP fields with associated values. The user can interactively compose filtering predicates using AND and OR to construct compound filters. The system converts interactively expressed search filters in this view into SQL commands, which are then shown in the main content panel. For reference, the SQL command corresponding to the filtering options selected in Figure 7 is SELECT LOG_TIMESTAMP, RAW_REQUEST, MODEL_LABEL, SNORT_LABEL FROM HTTPLOG_REQUEST_LABELED WHERE MODEL_LABEL > 0.70 AND RAW_REQUEST LIKE
'%"getRemoteAddr" : "81.174.251.27"%' AND RAW_REQUEST LIKE
'%"getRequestURI" : "%index.php"%' ORDER BY MODEL_LABEL. For more complex searches (e.g., those based on nested conditions), users can directly enter custom raw SQL queries in the search tab.

The data table shows a filtered list of labeled entries in the database. There are three columns that correspond to the "EntryID", "Predicted Label", and "Snort Label". The EntryIDs are the timestamps that correspond to unique HTTP requests. Importantly, each timestamp is a clickable link that generates a popup containing the raw HTTP field-value pairs in a human-readable format. Figure 8 displays a popup that results from clicking on the first EntryID from the data table in Figure 7. Note that the value of "getRemoteAddr" matches the input value of "81.174.251.27", and the value of "getRequestURI" matches the input value of "%index.php", in which % denotes a wildcard character that represents zero or more characters. The predicted label is the outputted probability from the previously described ML model that predicts anomalous Nectar Network requests. The output is consistent with the anomaly threshold option selected in the sidebar that corresponds to a minimum value of 0.7.

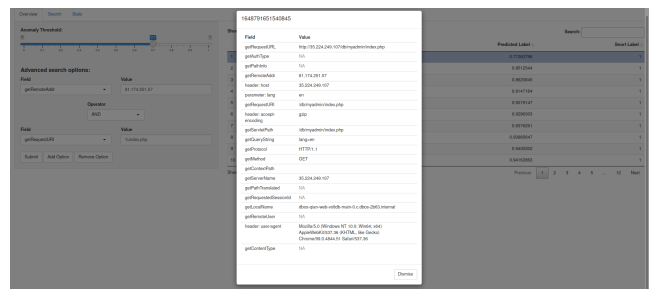

Figure 8: Popup displaying HTTP request information for a particular entry.

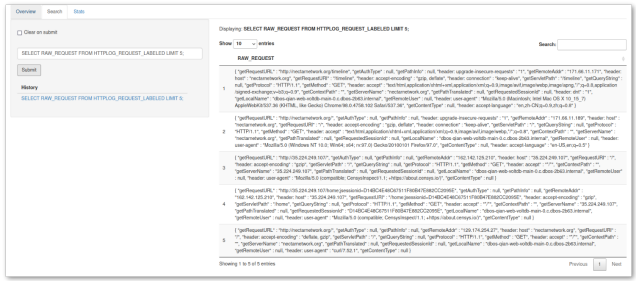

**Figure 9: Search page enables users to directly query the provenance database.**

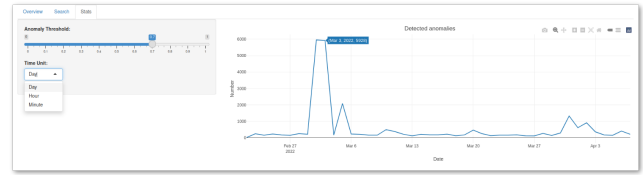

**Figure 10: Stats page enables users to easily explore patterns of anomalous activity across time.**

The Snort label, generated by running a public Snort ruleset on the data, is the ground truth used to train and evaluate the machine learning model. Notably, new entries will only have a predicted label with no available Snort label. By default, entries are sorted in ascending order by the predicted label, but there are ascending and descending sort options available for each column.

**Investigating Anomalies** The main function of the search page is to enable the user to directly query the underlying Vertica provenance database. This is done by inputting and executing arbitrary SQL queries, which gives the flexibility needed to handle complex investigations. An example SQL command has been executed and displayed in Figure 9. The left sidebar specifies user inputs and options, and the main content panel displays the result of the SQL query. The sidebar provides several features to facilitate the search process. The top checkbox dictates whether the user text input below should be cleared every time a new query is submitted. Importantly, queries are not sanitized or modified before submission. The web application operates under the assumption that it is being used by a security administrator or other party with intimate knowledge of the underlying Vertica provenance database schema. An additional quality of life feature is that queries can be submitted by either using the submit button or pressing the keyboard enter button. Once a query has been submitted and displayed, a link is generated under the history section with the exact text submitted. This allows the user to easily resubmit a previous query by simply clicking on the associated link. Note that the history only maintains the previous 10 submissions, and each submission in the history is unique. Repeatedly submitting the same query will simply result in that query staying at the top of history.

The main panel of the search page (Figure 9) consists of the table that results from the most recent query submitted by the user. Importantly, the text of the most recent query is displayed in bold above the table for ease of use. In the event of an invalid SQL query, the web application handles it gracefully by displaying a two-entry data table that consists of the problematic syntax as well as the original query string.

**Analyzing Historical Data** The stats page provides a historical view of anomalous behavior by visualizing aggregate anomalous activity over various timeframes and dates. The default settings and associated line plot are shown in Figure 10. The left sidebar specifies all available user options, and the main content panel displays the resulting line plot. There is no submit button as in the other pages since the line plot reacts automatically to changes or selections in the user options. This is simply because there are no text inputs in the stats user options that would cause unnecessary and excessive queries upon all textual changes whatsoever. Users can use the historical view to recognize peaks in the number of attacks that might point to a concerted attack effort or to understand long-term trends in their security. The sidebar consists of only two components: the anomaly threshold and the time unit. As in the overview page, the anomaly threshold dictates the minimum value at which predicted labels are considered to be truly anomalous. The time unit has three options: day, hour, and minute. The time unit indicates the granularity at which anomalies should be aggregated and displayed in the line plot.

The main view displays an interactive line plot with the selected user options. Users can select and magnify subsections of the line graph with brushing. Similarly, users can mouse hover the line graph data points and see the corresponding tuple of the exact date and number of detected anomalies in a tooltip. Overall, users can download the plot, zoom freely, pan, zoom in, zoom out, autoscale, reset axes, show the closest data on hover, and compare data on hover.

### 3.4 User Interface Evaluation

Throughout the development of the application, we actively sought out user feedback to improve the application design and better understand the needs of target users. We conducted a longitudinal study involving five industry professionals. We collected feedback through two consecutive studies, a formative (initial) study carried out at the beginning of the development and a summative (final) study at the end, using an identical protocol. The participants were given a brief explanation of the purpose of the web application as well as a live demonstration of its main features at the time. After each view was described and displayed in its entirety, each user was allowed to freely explore the view to their satisfaction. Afterward, they were given view-specific questions (Table 5) to evaluate the key aspects of views such as functionality, aesthetics, and ease of use. Our participants were asked to give numeric scores between 1 (strongly disagree) to 5 (strongly agree) for each question. They were also given an open-ended prompt at the end of each survey to provide general feedback and suggestions.

**Table 5: Survey questions. We elicited responses for twelve view-specific questions, four per view.**

| View (Page) | | Question |
|---|---|---|
| Overview | Q1 | *How well do you feel you can prioritize anomalies based on the available sidebar options?* |
| | Q2 | *Does the data table display relevant anomaly information clearly and effectively?* |
| | Q3 | *Would you consider the layout of the page to be well-organized and aesthetically pleasing?* |
| | Q4 | *Is it simple and easy to use the provided interface to produce your desired output?* |
| Search | Q5 | *How well do you feel you can investigate anomalies based on the available sidebar options?* |
| | Q6 | *Does the data table display relevant anomaly information clearly and effectively?* |
| | Q7 | *Would you consider the layout of the page to be well-organized and aesthetically pleasing?* |
| | Q8 | *Is it simple and easy to use the provided interface to produce your desired output?* |
| Stats | Q9 | *How well do you feel you can understand historical anomaly trends based on the available sidebar options?* |
| | Q10 | *Does the line plot display relevant anomaly information clearly and effectively?* |
| | Q11 | *Would you consider the layout of the page to be well-organized and aesthetically pleasing?* |
| | Q12 | *Is it simple and easy to use the provided interface to produce your desired output?* |

**Initial Study** The initial study was performed to gather formative feedback on the general functionality, design, and implementation of the system. Figure 11 shows the aggregate scores (green) given by the participants after the guided, live demonstration, and free exploration process of each page. Scores were generally above average, but there were some poorer ratings in the overview and search pages that warranted further development. The general feedback regarding both views was that the data table returned by queries was relatively cluttered and lacking in functionality. At the same time, the limited number of available side options in both pages made it difficult to truly prioritize or investigate anomalies in depth.

We implemented several changes based on the initial study to improve the user interface and general functionality of the application. In particular, we implemented a new data table rendering to support search options within the table itself as well as column-specific sorting options (ascending or descending order). At the same time, we removed the column showing the HTTP field-value pairs as plain text as another in the overview page. Instead, we used the timestamps as unique entry IDs with clickable links. When clicked on, these links produced popups that displayed HTTP field-value pairs in a standard table format, making it much more accessible. With respect to the overview sidebar options, the advanced search options were added to let users filter based on any of the fields in the HTTP requests. This enables users to quickly and efficiently dive into only the requests they're interested in. With respect to the search sidebar options, the unique history provided a much needed quality of life improvement that made investigations much simpler and easier. Users can now recall a previously executed query to see the result again or to make small edits to the query as they explore the data. Additionally, the line plot in the stats page was replaced and rendered using the Plotly package [32] rather than ggplot. This allowed for interactive features including panning, zooming, and brushing incorporated directly into the plot itself.

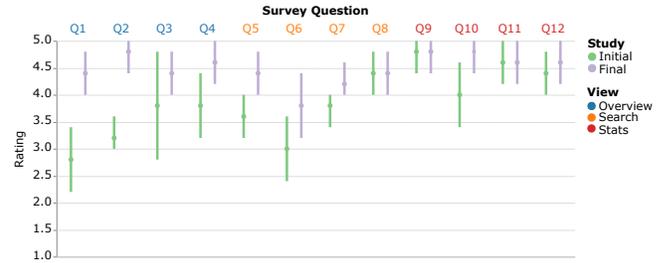

**Figure 11: User feedback from the initial and final studies. The final feedback is promising and represents a consistent improvement (higher means and lower variances) over the initial feedback across all survey questions.**

**Final Study** The summative study was performed after incorporating the user feedback from the initial study as we discussed above. Figure 11 shows aggregate responses (purple) obtained from our participants in the final study. Results represent a significant improvement over the initial evaluation, suggesting our revisions on the user interface and the underlying functionalities have been effective.

## 4 CONCLUSION

DBOS facilitates ML applications by supporting the following three principles.

**(1) A DBMS for ML applications is a very good idea.** The model and all ancillary data should be DBMS-based, for easy querying and versioning. The DBOS design highly encourages this mode of thinking.

**(2) Co-locating data and computation is a very good idea.** Whenever possible, DBOS co-locates computation and data. Anything coded in our programming environment [23] supports this, and many ML operations can be done this way, resulting in a significant performance speedup. Machine learning on specialized hardware cannot leverage co-location; however, DBOS performance using accelerators is similar to that of existing approaches.

**(3) Automatic provenance support is a very good idea.** Not only does this facilitate backing up an ML project so a different path forward can be tried, but also monitoring/real-time/security ML applications can run directly off the provenance data as we showed in this paper.

In summary, ML applications are data intensive and model evolution can be well supported by a versioning system. Such applications benefit from the DBOS architecture, and we expect these ideas will become more prevalent in the future.